\documentclass[letterpaper, 10 pt, journal, twoside]{IEEEtran}
\usepackage{cite}
\usepackage{amsmath,amssymb,amsfonts}
\usepackage{algorithmic}
\usepackage{graphicx}
\usepackage{textcomp}

\usepackage{tikz}
\usetikzlibrary{shapes,arrows}
\usepackage{nicefrac}
\usepackage{mathtools, cuted}
\usepackage{lipsum}
\usepackage{float}
\usepackage{afterpage}
\usepackage{placeins}
\usepackage{tabularx}
\usepackage{balance}
\usepackage{textcomp}%
\usepackage{booktabs}
\usepackage{multirow}
\usepackage{accents}
\usetikzlibrary{patterns,shapes,arrows}
\usetikzlibrary{positioning}
\usepackage{subfig}
\allowdisplaybreaks
\allowdisplaybreaks[1]

\tikzstyle{decision} = [diamond, draw, fill=blue!20,
    text width=7em, text badly centered, inner sep=0pt]
\tikzstyle{decision1} = [diamond, draw, fill=blue!20,
    text width=7em, text badly centered, inner sep=0pt, node distance = 2.5cm]
\tikzstyle{block} = [rectangle, draw, fill=blue!20,
    text width=13em, text centered, rounded corners, minimum height=3em]
\tikzstyle{block2} = [rectangle, draw, fill=blue!20,
    text width=5em, text centered, rounded corners, minimum height=3em, node distance = 4cm]
\tikzstyle{block3} = [rectangle, draw, fill=red!20,
    text width=5em, text centered, rounded corners, minimum height=3em, node distance = 4cm]
\tikzstyle{line} = [draw, very thick, color=black!50, -latex']
\tikzstyle{cloud} = [draw, ellipse,fill=red!20, node distance=2.5cm, minimum height=2em]

\begin{document}

\title{Reliable Solution to Dynamic Optimization Problems using Integrated Residual Regularized Direct Collocation}

\author{Yuanbo Nie and Eric C.\ Kerrigan \IEEEmembership{Senior Member, IEEE}
\thanks{Yuanbo Nie is with the School of Electrical \& Electronic Engineering, University of Sheffield, S1~3JD Sheffield, U.K.. {\tt\small y.nie@sheffield.ac.uk}}%
\thanks{Eric C. Kerrigan is with the Department of Electrical \& Electronic Engineering and Department of Aeronautics, Imperial College London, London SW7~2AZ, U.K., {\tt\small 
e.kerrigan@imperial.ac.uk}}%
}

\maketitle

\thispagestyle{empty}

\begin{abstract}
Direct collocation is a widely used method for solving dynamic optimization problems (DOPs), but its implementation simplicity and computational efficiency are limited for challenging problems like those involving singular arcs. In this paper, we introduce the direct transcription method of integrated residual regularized direct collocation (IRR-DC). This method enforces dynamic constraints through a combination of explicit constraints and penalty terms within discretized DOPs. This method retains the implementation simplicity of direct collocation while significantly improving both solution accuracy and efficiency, particularly for challenging problem types. Through the examples, we demonstrate that for difficult problems where traditional direct collocation results in excessive fluctuations or large errors between collocation points, IRR-DC effectively suppresses oscillations and yields solutions with greater accuracy (several magnitudes lower in various error metrics) compared to other regularization alternatives.
\end{abstract}

\begin{IEEEkeywords}
dynamic optimization, optimal control
\end{IEEEkeywords}

\section{Introduction} \label{sec:Intro}
\IEEEPARstart{O}{ptimization} problems involving dynamical systems are a crucial aspect of engineering, from optimal control to design optimizations. For practical dynamic optimization problems (DOPs), numerical approaches are commonly employed, with the direct transcription method being the most widely used. This method converts the infinite-dimensional problems into finite-dimensional discretised DOPs (DDOPs).

For short-horizon problems, e.g.\ those typically used in model predictive control (MPC), the transcription method of multiple shooting is a very popular choice~\cite{diehl2009efficient}. For longer horizon optimal control problems and DOPs with sophisticated constraints, the direct collocation (DC) method is generally considered to be preferable thanks to its well-rounded balance between computation complexity and solution accuracy~\cite{betts2010practical}. 

However, certain categories of DOPs, particularly those involving singular arcs and high-index differential-algebraic equations (DAEs), present challenges for DC. The development of integrated residual methods~\cite{MartinThesis, TACDAIR, neuenhofen2023numerical} offers an alternative approach that can address these cases with greater reliability. Nonetheless, this method is more flexible, hence typically requires higher expertise to configure the transcription process appropriately. As a result, DC remains the go-to numerical method for solving a general DOP with limited prior knowledge.

This paper explores how the core framework of DC can be maintained while incorporating insights and techniques from the development of integrated residual transcription methods, aiming to enhance its ability to address challenging problems, such as those involving singular arcs. We introduce this approach as the integrated residual regularized direct collocation (IRR-DC) in Section~\ref{sec:IRRDC}. In Section~\ref{sec: ExampleProblem}, we provide numerical examples to showcase the advantages of the proposed method.

\section{Numerical Solution of Dynamic Optimisation Problems}
\label{sec: OptimizationBasedControl}

Optimization-based control often requires the solution of  DOPs expressed in the general Bolza form:
\begin{subequations}
\label{eqn:cont_DOP}
\begin{equation}
    \min_{\substack{x(\cdot),u(\cdot)\ \\ t_0,t_f,p}} \label{eqn:DOPBolzaObjective}
    \begin{array}{l}
    V_M(x(t_0),x(t_f),u(t_0),u(t_f),t_0,t_f,p) +  \\ 
    \int_{t_0}^{t_f}\ell(\dot{x}(t),x(t),u(t),t,p) dt
    \end{array}
\end{equation}
subject to
\begin{align}
    f(\dot{x}(t),x(t),u(t),t,p) = 0,\ & \label{eqn:DEs}\\
    g(\dot{x}(t),x(t),\dot{u}(t),u(t),t,p) \leq 0,\  &\label{eqn:ineqs}\\
    b_E(x(t_0),x(t_f),u(t_0),u(t_f),t_0,t_f,p) = 0,\ & \label{eqn:terminal_eq}\\
    b_I(x(t_0),x(t_f),u(t_0),u(t_f),t_0,t_f,p)\leq 0,\ \label{eqn:terminal_ineq}&
\end{align}
\end{subequations} 
 with $x: \mathbb{R} \rightarrow \mathbb{R}^n$ is the state trajectory of the system, $u: \mathbb{R} \rightarrow \mathbb{R}^m$ is the control input trajectory,   $p \in \mathbb{R}^s$ are static parameters, $t_0 \in \mathbb{R}$ and $t_f \in \mathbb{R}$ are the initial and terminal time.  $V_M$ is the Mayer cost functional ($V_M$: $\mathbb{R}^n  \times \mathbb{R}^n \times \mathbb{R}^m \times \mathbb{R}^m \times \mathbb{R} \times \mathbb{R} \times \mathbb{R}^s \to \mathbb{R}$), $l$ is the Lagrange cost functional ($l:\mathbb{R}^n \times \mathbb{R}^n \times \mathbb{R}^m \times \mathbb{R} \times \mathbb{R}^s \to \mathbb{R}$), $f$ is the dynamic constraint ($f:\mathbb{R}^n \times \mathbb{R}^n \times \mathbb{R}^m \times \mathbb{R} \times \mathbb{R}^s \to \mathbb{R}^{n_f}$), $g$ is the path constraint ($g:\mathbb{R}^n \times  \mathbb{R}^n \times \mathbb{R}^m \times \mathbb{R}^m \times \mathbb{R} \times \mathbb{R}^s \to \mathbb{R}^{n_g}$), and $b_E$ and $b_I$ are the boundary constraints ($b_E:\mathbb{R}^n  \times \mathbb{R}^n \times \mathbb{R}^m \times \mathbb{R}^m \times \mathbb{R} \times \mathbb{R} \times \mathbb{R}^s \to \mathbb{R}^{n_{bE}}$, $b_I:\mathbb{R}^n  \times \mathbb{R}^n \times \mathbb{R}^m \times \mathbb{R}^m \times \mathbb{R} \times \mathbb{R} \times \mathbb{R}^s \to \mathbb{R}^{n_{bI}}$).

\subsection{Direct Transcription methods}
\label{sec: DirectTranscriptionMethod}
To yield a practical approach to  solve the infinite-dimensional continuous-time DOP numerically, direct transcription methods transcribe the DOP into finite-dimensional DDOPs, typically resulting in linear programming (LP), quadratic programming (QP) and nonlinear programming (NLP) problems. 

In this transcription process, the state and input trajectories~$x$ and $u$ are approximated by parameterised approximation functions~$\tilde{x}$ and $\tilde{u}$, respectively. A popular choice is to define~$\tilde{x}$ and $\tilde{u}$ as piece-wise functions based on a time mesh with~$K$ intervals, with the $k^\text{th}$ interval denoted as $\mathbb{T}_k:=[s_k,s_{k+1}]$. $s_k$ represent the interface between intervals of the time mesh $t_0=s_1 <\dots< s_{K+1}=t_f$, also known as \emph{major node} locations.

Under this setting, the trajectory for state variables inside each interval $k$ can be approximated as
\begin{equation}
\label{eqn: LGRStateApproximation}
x^{(k)}(t) \approx \tilde{x}^{(k)}(t) := \sum_{i=1}^{N^{(k)}}\chi_i^{(k)}\beta_{i}^{(k)}(t),
\end{equation}
with $\beta_{i}^{(k)}(\cdot)$ a basis function and $\chi_i^{(k)}$ the corresponding coefficient. For convenience, practical implementation could choose interpolating polynomials as the basis functions, making the coefficients correspond to $N^{(k)}$ points on the polynomial function $\tilde{x}^{(k)}$. A useful property of such choices is that there exists time instances $s_k \le d_1^{(k)} <\dots< d_{N^{(k)}}^{(k)} \le s_{k+1} $ where $\chi_i^{(k)}=\tilde{x}^{(k)}(d_i^{(k)}) \in \mathbb{R}^{n}$ for all $i\in\mathbb{I}_{N^{(k)}}$. In this case, the unknown coefficients $\chi_i^{(k)}$ are sampled points of the approximate state trajectory known as the \emph{parameterized states}, and the corresponding time instances $d_i^{(k)}$ are named the \emph{data points}. The parameterization of the input can be done similarly with $\upsilon_i^{(k)}=\tilde{u}^{(k)}\left(d_i^{(k)}\right) \in \mathbb{R}^{m}$.

With $\chi: = [\chi^{(1)},\dots, \chi^{(K)}]^{\top}$, $\chi^{(k)}=[\chi_1^{(k)},\dots, \chi_{N^{(K)}}^{(k)}]^{\top}$, $\upsilon: = [\upsilon^{(1)},\dots, \upsilon^{(K)}]^{\top}$, $\upsilon^{(k)}=[\upsilon_1^{(k)},\dots, \upsilon_{N^{(K)}}^{(k)}]^{\top}$, a general formulation for the DDOP is:
\begin{subequations}
\label{eqn:DDOPProblem}
\begin{multline}
\label{eqn:DDOPObjective}
\min_{\chi,\upsilon,t_0,t_f,p}  V_M\left(\chi_1^{(1)},\chi_{N^{(K)}}^{(K)},t_0,t_f,p\right)\\
+\sum_{k=1}^{K}\sum_{i=1}^{Q^{(k)}} w_i^{(k)} l\left(\tilde{x}^{(k)}(q_i^{(k)}),\tilde{u}^{(k)}(q_i^{(k)}),t_0,t_f,p\right)
\end{multline}
subject to, for all $k\in\mathbb{I}_K$, 
\begin{align}
\label{eqn:DDOPEQNConstraint}
\psi^{(k)} \left(\chi^{(k)},\upsilon^{(k)},t_0,t_f,p\right) =  0,&\\
\label{eqn:DDOPIEQNConstraint}
\gamma^{(k)} \left(\chi^{(k)},\upsilon^{(k)},t_0,t_f,p\right) \le  0,&
\end{align}
and for some $k_i\in\mathbb{I}_K$ and $k_j\in\mathbb{I}_K$,
\begin{align}
\label{eqn:DDOPEQNBoundaryConstraint}
\phi_E\left(\chi_1^{(k_i)},\chi_{N^{(K)}}^{(k_j)},\upsilon_1^{(k_i)},\upsilon_{N^{(K)}}^{(k_i)},t_0,t_f,p\right) = 0,&\\
\label{eqn:DDOPIEQNBoundaryConstraint}
\phi_I\left(\chi_1^{(k_i)},\chi_{N^{(K)}}^{(k_j)},\upsilon_1^{(k_i)},\upsilon_{N^{(K)}}^{(k_i)},t_0,t_f,p\right) \le 0.&
\end{align}
\end{subequations}
To approximate the integral of the Lagrange cost inside an interval $k$, numerical integration with a $Q^{(k)}$-point quadrature rule is used, with \emph{quadrature weights} $w_i^{(k)}$ and \emph{quadrature abscissae} $q_i^{(k)}$, for all $i\in\mathbb{I}_{Q^{(k)}}$. Constraints~\eqref{eqn:DDOPEQNBoundaryConstraint}--\eqref{eqn:DDOPIEQNBoundaryConstraint} can be used to enforce the boundary constraints~\eqref{eqn:terminal_eq}--\eqref{eqn:terminal_ineq}, as well as any continuity constraints for state and/or input trajectories.

\subsection{Handing of dynamic constrains}
The handling of the dynamic constraint~\eqref{eqn:DEs} is where different direct transcription methods differ. While the direct multiple shooting method utilizes external numerical solvers for differential equations to forward integrate the dynamic equations (e.g.\ Runge-Kutta integrators), the direct collocation and integrated residual transcription methods directly embed the (approximate) satisfaction of the dynamic constraints inside the formulation of the DDOP.

\subsubsection{Implementing dynamic relationships as equality constraints in the DDOP}
From the weighted residuals method~\cite[Sect.~5.8]{rao2010finite} for the numerical solution of differential equations, the satisfaction of the dynamic equations requires 
\begin{equation}
\label{eqn: WRMWeakForm}
    \int_{\mathbb{T}_k} \varpi^{(k)}(t)r_{\xi}^{(k)}(t)\: dt =0, \quad \forall \xi\in\mathbb{I}_{n_f},
\end{equation}
to hold. $\varpi^{(k)}$ is commonly referred to as a \emph{test function} as the relationship tests the value of the local residual, defined as
\begin{equation}
\label{eqn: LocalResidual}
    r^{(k)}(t) := f(\dot{\tilde{x}}^{(k)}(t),\tilde{x}^{(k)}(t),\tilde{u}^{(k)}(t),t,p).
\end{equation}
In the limit case where~\eqref{eqn: WRMWeakForm} holds for all test functions taken from a suitably-defined set of functions, the residual and hence the error will diminish to zero. In practice, the aim is to choose finite sets of test functions that would approximately solve the differential equation, with popular designs including the Galerkin, collocation, least-squares methods and the method of moments. Those different choices lead to different accuracy and computational complexity characteristics, making appropriate choices problem-specific.

The collocation method uses Dirac delta functions as test functions in~\eqref{eqn: WRMWeakForm}. With the isolation property of Dirac delta functions~\cite{TACDAIR}, \eqref{eqn: WRMWeakForm} will become a system of equations that requires the local residual~\eqref{eqn: LocalResidual} to be zero at the center of the Dirac delta functions, also known as \emph{collocation points}. 

In the direct collocation transcription method, the above-mentioned equation system are implemented in~\eqref{eqn:DDOPEQNConstraint} as 
\begin{multline}
f\Big(\dot{\tilde{x}}^{(k)}(d_i^{(k)}),\tilde{x}^{(k)}(d_i^{(k)}),\tilde{u}^{(k)}(d_i^{(k)}),d_i^{(k)},p\Big)=0, \\ \text{for all } i\in\mathbb{I}_{N^{(k)}}, \,k\in\mathbb{I}_K.
\end{multline}
Such an implementation can be made to be computationally efficient, and this aspect contributes to the popularity of the direct collocation approach in developing general-purpose DOP solvers: the benefit of using a more sophisticated setting, e.g.\ the Galerkin method, can only be exploited if the nature of the dynamical system is known beforehand. 

\subsubsection{Implementing dynamic relationships as inequality constraints or as penalty terms in the DDOP}
The drawback of only forcing the residual error to be zero at the collocation points is that large errors could still arise in-between the collocation points. When the trajectories are forward integrated, the errors are accumulated. This accumulation leads to constraint violations and significant deterioration in solution optimality should the numerical DOP solution be implemented.


The integrated residual direct transcription method addresses the challenge by enforcing the dynamic equations through the integrated form of the residuals~\eqref{eqn: LocalResidual}, namely
\begin{equation}
\label{eqn: ResidualMinimizationOrgr}
\hat{R}^{(k)}(\tilde{x}^{(k)}, \tilde{u}^{(k)}, t_0, t_f, p):= \int_{\mathbb{T}_k}  \|\alpha \circ r^{(k)}(t)\|^2_{2}\: dt.
\end{equation}
with $\alpha \in \mathbb{R}^{n_f}$ being additional weighting parameters, e.g.\ to account for the interval length $\Delta t=s_{k+1}-s_k$ and/or differences in the numerical range of variables and constraints. In practice, this integration could be approximated by a quadrature rule similar to the discrete approximation of the Lagrange cost in~\eqref{eqn:DDOPObjective}, as
\begin{equation}
\label{eqn: ResidualMinimizationDiscreteQuadrature}
\mathcal{R}\left(\chi,\upsilon,t_0,t_f,p\right):=\sum_{k=1}^{K}\sum_{i=1}^{Q^{(k)}}w_{i}^{(k)}\begin{Vmatrix}\alpha \circ r(q_i^{(k)})\end{Vmatrix}^2_{2}.
\end{equation}

For commonly used state and input trajectory parameterisations, approximation errors are inevitable~\cite{TACDAIR}. Hence, a DDOP with $\mathcal{R}$ forced to zero with an arbitrary design of the quadrature rule may not always be feasible except in a few special cases; one such special case is to select a quadrature rule such that the quadrature abscissae $q_i^{(k)}$ match the data points $d_i^{(k)}$, which recovers the direct collocation scheme~\cite{TACDAIR}. 

In other words, with a quadrature rule of sufficiently high order to capture the error behavior inside the interval, $\mathcal{R}$ cannot be reduced to zero in general. Instead, $\mathcal{R}$ needs to be minimized directly in the objective, or implemented as inequality constraints in the DDOP. 

The quadrature penalty method (QPM)~\cite{MartinThesis,neuenhofen2023numerical,9303897} minimizes the integrated residuals via an additional penalty term, with the DDOP being
\begin{equation}
\label{eqn:QPMObjective}
\min_{\chi,\upsilon,p,t_0,t_f}  J_h\left(\chi,\upsilon,t_0,t_f,p\right) + \frac{1}{2 \rho}\mathcal{R}\left(\chi,\upsilon,t_0,t_f,p\right)
\end{equation}
subject to~\eqref{eqn:DDOPIEQNConstraint}, \eqref{eqn:DDOPEQNBoundaryConstraint} and~\eqref{eqn:DDOPIEQNBoundaryConstraint}. $J_h \in \mathbb{R}$ is the original objective as in~\eqref{eqn:DDOPObjective} and $\rho>0$ is the regularization weight of the penalty term. 

In contrast, the direct alternating integrated residuals (DAIR) approach~\cite{TACDAIR} solves the problem by alternating between the minimization of the integrated residuals and the minimization of the original objective. The DAIR residual minimization problem is formulated as
\begin{subequations}
\begin{equation}
\label{eqn:DAIRRObjective}
\min_{\chi,\upsilon,p,t_0,t_f}  \mathcal{R}\left(\chi,\upsilon,t_0,t_f,p\right)
\end{equation}
subject to~\eqref{eqn:DDOPIEQNConstraint}, for all $k\in\mathbb{I}_K$, and~\eqref{eqn:DDOPEQNBoundaryConstraint} and~\eqref{eqn:DDOPIEQNBoundaryConstraint} for some $k_i\in\mathbb{I}_K$ and $k_j\in\mathbb{I}_K$, and optionally inequality constraints 
    \begin{align}
        J_h\left(\chi,\upsilon,t_0,t_f,p\right) & \le J_c,\\
        \label{eqn:DAIRRResidualConstraint}
        \sum_{k=1}^{K}\sum_{i=1}^{Q^{(k)}}w_{i}^{(k)}\left(\alpha_{\xi} r_{\xi}(q_i^{(k)})\right)^2 & \le \Lambda_{\xi}, \quad \forall \xi\in\mathbb{I}_{n_f}
    \end{align}
with $J_c \in \mathbb{R}$ an upper bound on the original objective, and $\Lambda \in \mathbb{R}^{n_f}$ upper bounds for the mean integrated residual squared (MIRS) error of each individual dynamic equation.
\end{subequations}

The DAIR cost minimization problem is 
\begin{equation}
\label{eqn:DAIRCObjective}
\min_{\chi,\upsilon,p,t_0,t_f}  J_h\left(\chi,\upsilon,t_0,t_f,p\right)
\end{equation}
subject to~\eqref{eqn:DDOPIEQNConstraint},~\eqref{eqn:DDOPEQNBoundaryConstraint},~\eqref{eqn:DDOPIEQNBoundaryConstraint} and~\eqref{eqn:DAIRRResidualConstraint}. DAIR solves those two problems either till convergence or to an early termination criterion with details outlined in~\cite{TACDAIR}. 

Working with the residual error in the integrated form can bring a number of benefits, namely 
\begin{itemize}
    \item solutions of higher accuracy compared to DC can be obtained for a given discretisation mesh (with respect to commonly used error criteria),
    \item additional flexibility is available with regards the trade-off between solution accuracy and optimality,
    \item provides more reliable numerical solutions for challenging problems with singular arcs and high-index differential algebraic equations.
\end{itemize}

\subsection{Error measures in direct transcription methods}
Convergence of the DDOP~\eqref{eqn:DDOPProblem} to a solution does not fully resolve the numerical solution of the DOP~\eqref{eqn:cont_DOP}. This is because, regardless of the chosen direct transcription method, the initial discretization mesh may not be suitable for achieving the required accuracy. As a result, mesh refinement (MR) may be necessary as a subsequent step, based on an assessment of the errors in the approximate solution obtained from the DDOP.

Commonly used error measures include
\begin{align*}
    \Big|J_h\left(\chi,\upsilon,t_0,t_f,p\right) - J^{\ast}\Big| \leq \delta\,,&\\
    \int_{\mathbb{T}_k} 	|r_{\xi}^{(k)}(t)|\: dt \leq \eta_{\xi}, \quad \forall k \in \mathbb{I}_{N^{(k)}}, \xi\in\mathbb{I}_{n_f},&\\
    g(\dot{x}(t),x(t),\dot{u}(t),u(t),p,t) \leq \sigma, \quad \forall t \in [t_0,t_f]\,.&
\end{align*}
The first equation is the \emph{optimality gap}, with $J^{\ast}$ the optimal objective for the original DOP. The second measure is \emph{the absolute local (AL) error}, reflecting the violation of the dynamic equations locally inside a mesh interval. A scaled version of this is known as the \emph{the relative local (RL) error}~\cite{betts2010practical}. The last measure regarding the inequality path constraints is the \emph{constraint violation (CV) error}. 

Except for a few test problems where the optimal solution can be analytically derived, the exact solutions are not obtainable for the majority of the practical problems; hence, the optimality gap is often only used for benchmarking purposes. In contrast, the measure of absolute local error and constraint violation error provide valuable information regarding the necessary modifications to the discretisation mesh for solution improvements. Therefore, most MR schemes use one or both criteria for mesh design iterations. 

\subsection{Characteristics of Direct Collocation and Integrated Residual Approaches}
By forcing the residual to be zero only at collocation points, DC offers a computationally efficient transcription of the DOP into the DDOP and a numerical solution of the DDOP. To address the challenge of errors arising in-between collocation points, posterior error analysis and MR is often considered as an integral part of the DC scheme to yield solutions of good accuracy. This often works well in practice except for a number of challenging cases.

For DOPs with singular arcs and high-index DAEs, excessive fluctuations can be present in the numerical solution. With a denser mesh, the frequency of the fluctuations can increase but the amplitude of the fluctuations may not necessarily reduce. As a result, mesh refinement can be inefficient and ineffective in reducing the errors in the solution. To suppress the fluctuations in practical implementations, it is common to introduce regularization terms in the objective to penalize control actions~\cite{BBSOCP} (with $\int_{t_0}^{t_f} u^{\top}(t)Ru(t)\,dt$ and $R$ a weighting matrix, referred to later as the input regularization) or to penalize the rate of change of the control (with $\int_{t_0}^{t_f} \dot{u}^{\top}(t)R\dot{u}(t)\,dt$, referred to later as the input rate regularization). For DOPs with a large number of input variables and complex structures, managing the regularization to limit its impact on solution optimality can be quite difficult. 

On the other hand, the integrated residual approach, as a generalization of the collocation method, offers greater flexibility in balancing solution accuracy and optimality on a given discretization mesh~\cite{TACDAIR}. It is also more reliable when handling challenging problems where DC encounters difficulties~\cite{MartinThesis}. However, this flexibility can be a double-edged sword. While it allows for more accurate solutions of the DOP on coarser meshes, potentially saving computation time, it also requires additional expertise and care.  For instance, the scaling of the variables and dynamic constraints, as well as the choice of trade-offs in different aspects of the numerical optimal solution (e.g. optimality vs. accuracy), can significantly affect the solution process. Moreover, efficient and reliable implementations of QPM require specialized DDOP solver designs, and both QPM and DAIR necessitate additional configuration parameters in comparison to DC.

\section{Integrated Residual Regularized Direct Collocation}
\label{sec:IRRDC}
The development of IRR-DC aims to bring the benefit of integrated residual approaches in handling challenging problems to the DC framework, while maintaining DC's ease of implementation and computational efficiency as much as possible. 

The DDOP formulation arising from IRR-DC is as follows:
\vspace{-6mm}
\begin{subequations}
\label{eqn:IRRDCProblem}
\begin{multline}
\label{eqn:IRRDCObjective}
\min_{\chi,\upsilon,p,t_0,t_f}  J_h\left(\chi,\upsilon,t_0,t_f,p\right) + \frac{1}{2 \rho}\mathcal{R}\left(\chi,\upsilon,t_0,t_f,p\right)
\end{multline}
subject to, for all $k\in\mathbb{I}_K$ and $i\in\mathbb{I}_{N^{(k)}}$, 
\begin{align}
\label{eqn:IRRDCEQNConstraint}
f\Big(\dot{\tilde{x}}^{(k)}(d_i^{(k)}),\tilde{x}^{(k)}(d_i^{(k)}),\tilde{u}^{(k)}(d_i^{(k)}),d_i^{(k)},p\Big)=0,&\\
\label{eqn:IRRDCIEQNConstraint}
\gamma^{(k)} \left(\chi^{(k)},\upsilon^{(k)},t_0,t_f,p\right) \le  0,
&\end{align}
and for some $k_i\in\mathbb{I}_K$ and $k_j\in\mathbb{I}_K$,
\begin{align}
\label{eqn:IRRDCEQNBoundaryConstraint}
\phi_E\left(\chi_1^{(k_i)},\chi_{N^{(K)}}^{(k_j)},\upsilon_1^{(k_i)},\upsilon_{N^{(K)}}^{(k_i)},t_0,t_f,p\right) = 0,&\\
\label{eqn:IRRDCIEQNBoundaryConstraint}
\phi_I\left(\chi_1^{(k_i)},\chi_{N^{(K)}}^{(k_j)},\upsilon_1^{(k_i)},\upsilon_{N^{(K)}}^{(k_i)},t_0,t_f,p\right) \le 0.&
\end{align}

For DOPs with consistent overdetermined constraints~\cite{neuenhofen2023numerical, TACDAIR}, \eqref{eqn:IRRDCEQNConstraint} may make the DDOP infeasible leading to convergence challenges. Under the IRR-DC framework with the integrated residuals directly penalized in the objective, it is also possible to relax~\eqref{eqn:IRRDCEQNConstraint} to
\begin{equation}
\label{eqn:IRRDCIEQNDynConstraint}
-\epsilon \leq f\Big(\dot{\tilde{x}}^{(k)}(d_i^{(k)}),\tilde{x}^{(k)}(d_i^{(k)}),\tilde{u}^{(k)}(d_i^{(k)}),d_i^{(k)},p\Big)\le\epsilon,\\
\end{equation}
\end{subequations}
instead, with $\epsilon$ a vector of suitably chosen small constants.

The choice of $\rho$ in IRR-DC can be seen as a mechanism for balancing solution accuracy and optimality for a given discretization mesh design. A large value of $\rho$ drives the solution of~\eqref{eqn:IRRDCProblem} towards the DC solution, prioritizing the reduction of the original objective. However, as later shown in Figure~\ref{fig:GoddardRocketObj} with the example problem, a low objective value reported by the DDOP solver may be misleading. As $\rho$ decreases, the method can automatically converge to a more accurate solution when multiple solutions exist with minimal or negligible differences in nominal cost. Further reduction of $\rho$ leads to a solution that is heavily biased towards higher accuracy, albeit at the cost of a higher objective value.

The practical advantage of IRR-DC is that, with~\eqref{eqn:IRRDCEQNConstraint} or~\eqref{eqn:IRRDCIEQNDynConstraint}, the need for carefully selecting configuration parameters (e.g.\ for $\rho$) and scaling parameters (e.g.\ for dynamic constraints) is significantly reduced compared to QPM. This allows the problem to be directly solved using most off-the-shelf DDOP solvers, in contrast to QPM, which prefers a tailored DDOP solver. Additionally, compared to the alternating iterative process of DAIR, IRR-DC is easier to implement, demands less prior knowledge and requires fewer configuration parameters.

\section{Example Problems}
\label{sec: ExampleProblem}
Here, we present two example problems to demonstrate the main advantages of the IRR-DC. Both DOPs are transcribed using the optimal control software \texttt{ICLOCS2}~\cite{ICLOCS2}, and numerically solved to a tolerance of $10^{-9}$ with NLP solver \texttt{IPOPT}~\cite{wachter2006implementation} (version 3.12.9). 

\FloatBarrier

\subsection{Singular Control Example: Goddard Rocket}

The Goddard rocket problem~\cite{goddard1920method} is a frequently used example for the analysis of optimal control problems with singular arcs. The problem aims to maximize the highest altitude reachable by a rocket using a fixed amount of propellant. Depending on the fidelity of the modeling of the atmospheric drag, different solution structures have been identified for the optimal control
input. When neglecting or considering linear drag only, the solution is shown to be  \textit{bang-bang}, i.e.\ to exhaust all propellant with maximum thrust  at launch and initial ascent, and then coasting to the highest point. With a quadratic drag model commonly used in subsonic flights~\cite{tsien1951optimum}, the optimal solution structure changes to \textit{bang-singular-bang} with an intermediate low thrust profile. 

\subsubsection{Suppression of singular arc fluctuations} In this example, we implement the Goddard rocket problem as described in \cite[Ex.\ 4.9]{betts2010practical}. Using DC, it is known for the solution to be oscillatory on the singular arc if no special treatment is implemented. To remove the singular arc oscillations, a multiphase formulation is typically used with additional constraints known as singular arc conditions imposed specifically for the second phase, which corresponds to the one with singular control. 

\begin{figure}[t]
    \centering
  \subfloat[State trajectories\label{fig:GoddardRocketStates}]{%
       \includegraphics[width=1\linewidth]{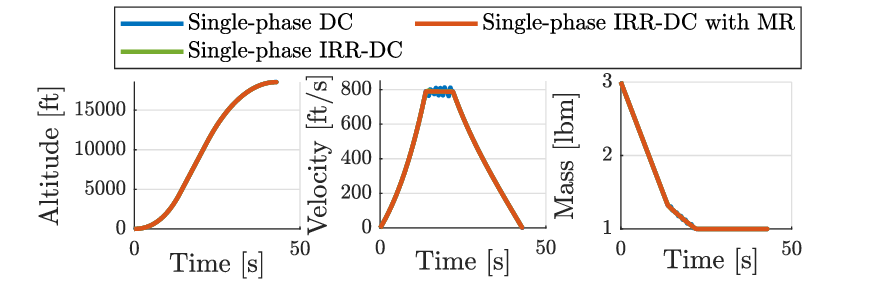}}
    \hfill
  \subfloat[Input trajectories\label{fig:GoddardRocketInput}]{%
        \includegraphics[width=1\linewidth]{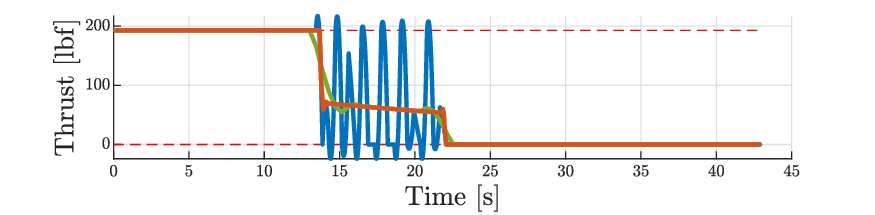}}
  \caption{Numerical solution to the Goddard rocket problem (Hermite-Simpson discretisation)}
  \label{fig:GoddardRocketSolution} 
\end{figure}

In~\cite{TACDAIR}, the ability of the integrated residual method of DAIR to alleviate the oscillations on the singular arc has been demonstrated on a fixed equidistant discretization mesh. The IRR-DC method yields similar improvements to the results: the large fluctuations on the singular arc have been suppressed  (Figure~\ref{fig:GoddardRocketSolution}), obtaining solutions of much higher accuracy in all measures (Table~\ref{tab: ErrorGoddard}). 

\begin{table}[t]
  \small
	\begin{center}
	\caption{Comparison of error measures}
	\label{tab: ErrorGoddard}
		\begin{tabular}{c|c|c|c|c}
		\hline  & \textbf{Max. AL} &\textbf{Max. RL} & \textbf{Max. CV}  & \textbf{Max. MIRNS}\\
          & \textbf{Error} & \textbf{Error} & \textbf{Error}  & \textbf{Error}\\
		\hline DC & 3.6e-01 & 1.1e-04 & 24.1 & 1.4e-03\\
            \hline IRR-DC & 2.6e-03 & 2.9e-06 & 3.6e-02 & 6.5e-08\\
		\hline
		\end{tabular} 
	\end{center}
\end{table}

\subsubsection{Mesh refinement performance}
Singular control problems are know to lead to challenges in MR: the convergence of MR iterations is often slow or unachievable. Table~\ref{tab: MRGoddard} demonstrates this: as the mesh gets denser, the maximum CV error fails to decrease under a standard DC scheme. This constraint violation is caused by the violation of the input bounds in-between collocation points on the singular arc with large fluctuations. With the singular arc fluctuations suppressed under IRR-DC, the MR process converges quickly with a considerably smaller final mesh size (also visualized in Figure~\ref{fig:GoddardRocketSolution}). 

\begin{table}[tb]
  \small
	\begin{center}
	\caption{Comparison of the mesh refinement history}
	\label{tab: MRGoddard}
		\begin{tabular}{c|c|c|c|c}
        \hline \multicolumn{5}{c}{DC}\\
		\hline \textbf{MR} & \textbf{Mesh} &\textbf{Max. AL } & \textbf{Max. CV}  & \textbf{NLP}\\
         \textbf{Iteration} & \textbf{Size} & \textbf{Error} & \textbf{Error}  & \textbf{Time}\\
		\hline 1 & 100 & 3.6e-01 & 24.1  & 1.5 s\\
		\hline 2 & 250 & 3.0e-01 & 24.1  & 3.9 s\\
		\hline 3 & 1171 & 9.3e-03 & 24.1  & 26.7 s\\
        \hline 4 & 2336 & 5.1e-03 & 24.1  & 44.9 s\\
        \hline 5 & 2555 & 1.5e-04 & 24.1  & 26.7 s\\
        \hline 6 & 2586 & 6.7e-05 & 24.1  & 41.5 s\\
        \hline 7 & 2624 & 6.7e-05 & 24.1  & 42.1 s\\
        \hline 8 & 2649 & 6.7e-05 & 24.1  & 44.4 s\\
		\hline
        \hline \multicolumn{5}{c}{IRR-DC}\\
		\hline \textbf{MR} & \textbf{Mesh} &\textbf{Max. AL} & \textbf{Max. CV}   & \textbf{NLP}\\
         \textbf{Iteration} & \textbf{Size} & \textbf{Error} & \textbf{Error}   & \textbf{Time}\\
		\hline 1 & 100 & 2.6e-03 & 3.6e-02  & 2.7 s\\
		\hline 2 & 138 & 5.6e-05 & 3.3e-03  & 4.5 s\\
		\hline 3 & 139 & 5.6e-05 & 6.4e-04  & 7.1 s\\
        \hline
		\end{tabular} 
	\end{center}
\end{table}

\subsubsection{Comparison with other commonly used regularization mechanisms}
\begin{figure}[t]
    \centering
  \subfloat[from DDOP solution\label{fig:GoddardRocketDOPObj}]{%
       \includegraphics[width=1\linewidth]{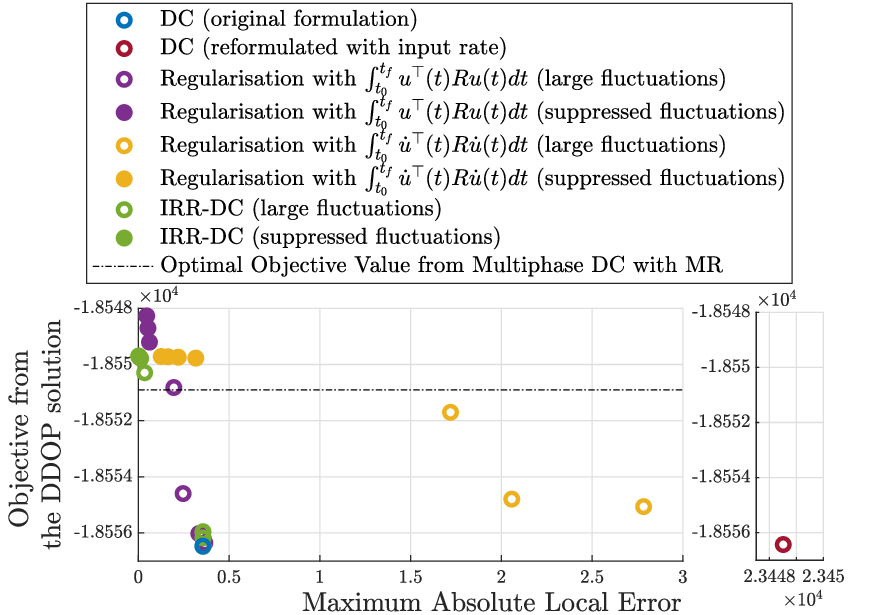}}
    \hfill
  \subfloat[from simulation\label{fig:GoddardRocketSIMObj}]{%
        \includegraphics[width=1\linewidth]{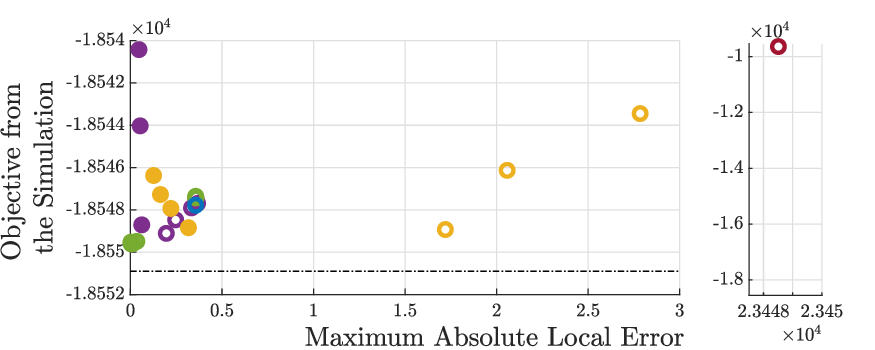}}
  \caption{Comparison of different regularization schemes for the suppression of singular arc fluctuations. Value for the dot-dashed line is from~\cite[Ex.\ 4.9]{betts2010practical}. }
  \label{fig:GoddardRocketObj} 
\end{figure}

Figure~\ref{fig:GoddardRocketDOPObj} compares the DDOP solution of IRR-DC solution against other regularization mechanisms. It can been seen in the figure that many of the solutions with singular arc fluctuations report lower objective values in comparison to that of a high accuracy solution~\cite[Ex.\ 4.9]{betts2010practical}. This highlights an important point of caution when comparing numerical DOP solutions:  the lower objective values obtained by solutions with higher errors may be misleading.




For a proper comparison, we obtained the actual objective by implementing the numerical DOP solutions through high-accuracy simulations instead and present the results in Figure~\ref{fig:GoddardRocketSIMObj}. 
All regularization schemes demonstrate the following trade-off behaviors:
\begin{itemize}
    \item with a small regularization weight (i.e.\ when 
$R$ is small and $\rho$ is large), large fluctuations may occur in the numerical solution, reducing its accuracy. Such solutions will result in significantly higher objective values when implemented, compared to those reported by the DDOP solver, making them highly undesirable.
    \item as the regularization weight increases (i.e.\ when $R$ increases and $\rho$ decreases), the fluctuations will be suppressed leading to an increase in solution accuracy. Such solutions are the desirable ones with low objective values that are realistically achievable.  
    \item with a further increase in regularization weight, the bias between the original objective and the regularization cost shifts. The solutions will become increasingly suboptimal with a higher nominal cost~\eqref{eqn:DOPBolzaObjective}. 
\end{itemize}


Next, we examine the differences in the solutions across various regularization schemes. Firstly, when optimizing with the original formulation, the inclusion of input regularization or integrated residual regularization both results in lower errors. In contrast, enabling input rate regularization --- by reformulating the dynamic equations (with the original input as a new state variable and the rate of change of the original input as the new input variable, as discussed in~\cite[Sect. IV.A]{TACRateConstraint}) --- can lead to further deterioration in solution accuracy, as shown in Figure~\ref{fig:GoddardRocketSIMObj}. Consequently, even with the penalty on input rate, the solution may still be less accurate compared to the DC solution using the original formulation. Therefore, additional care is needed when applying input rate regularization.

Secondly, we compare the solution optimality in the multi-objective sense, examining the trade-offs between solution optimality and solution accuracy, with a particular focus on solutions where singular arc fluctuations are suppressed. It is observed that the solution obtained by IRR-DC is the preferred one, as it simultaneously achieves the lowest objective and the lowest error compared to the other regularization alternatives. This is because IRR-DC penalizes the integrated residual error, rather than the inputs. Hence, IRR-DC will not discriminate against large inputs or large input rates that may be required in the optimal solution, as long as the errors are low.  

\subsection{Algebraic Constraint Example: Ventilator Control}
\label{sec: AircraftExample}
In the second example, we focus on demonstrating the benefit of the IRR-DC in handling problems with additional algebraic constraints, and on offering insights regarding the role of the regularization weights in IRR-DC. The problem presented is the ventilator control problem introduced in~\cite{Ventilator2020}. 

The problem formulation requires the enforcement of the following algebraic constraints:
\begin{equation*}
     R^I_{p}i_p(t)+a_{I,p}R_\delta i_p(t) = V_I(t)-v_p(t), \quad \forall p\in \mathbb{P}
\end{equation*}
with optimization decision variables $v_p(t)$ as the state, $i_p(t)$ and $V_I(t)$ as inputs, and $a_{I,p}$ as a static parameter. Additionally $R^I_{p}$ and $R_\delta$ are constants. 

\begin{figure}[tb]
\begin{center}
\includegraphics[width=0.95\columnwidth]{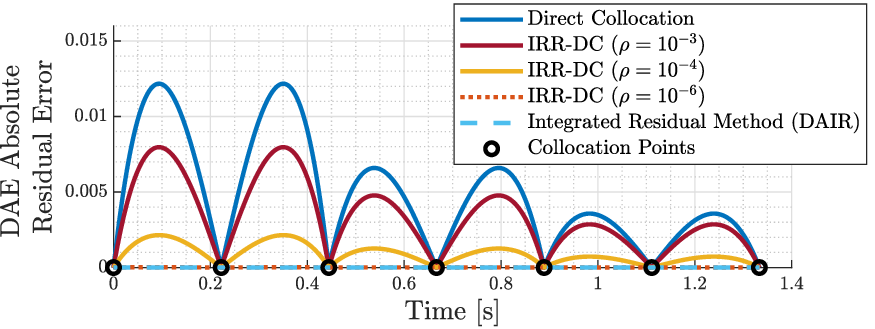}    
\caption{Comparison of residual error for the algebraic constraint in the ventilator control problem} 
\label{fig:ResidualErrorComparison}
\end{center}
\end{figure}

The integrated residual method's effectiveness in enhancing the precision of algebraic constraints with a coarse discretization mesh was shown in~\cite{Ventilator2020}. Figure~\ref{fig:ResidualErrorComparison} reproduces these solutions together with the solution with IRR-DC under different choices of $\rho$. With a decreasing $\rho$, the IRR-DC solution improves compared to the DC solution and tends to the integrated residual solution with DAIR. 

\section{Conclusions}
\label{sec: conlclusions}

Although direct collocation is a widely used direct transcription method for solving dynamic optimization problems, it is known to face difficulties with challenging problems, such as those involving singular arcs. The integrated residuals approach has shown greater capability in addressing such issues; however, its flexibility requires significant expertise to properly configure the transcription process. With the proposed integrated residual regularized direct collocation (IRR-DC), the dynamic constraints are incorporated as a penalty term in the objective function using an integral formulation, and as equality or inequality constraints for the residual error at the collocation points.

Our experience suggests that transcription with IRR-DC provides an easy-to-configure framework that retains the simplicity and computational efficiency of direct collocation, while greatly improving solution accuracy and reliability for challenging problems.

\bibliography{main} 

\begin{thebibliography}{10}

\bibitem{diehl2009efficient}
M.~Diehl, H.~J. Ferreau, and N.~Haverbeke, ``Efficient numerical methods for
  nonlinear {MPC} and moving horizon estimation,'' {\em Nonlinear model
  predictive control: towards new challenging applications}, pp.~391--417,
  2009.

\bibitem{betts2010practical}
J.~T. Betts, {\em Practical Methods for Optimal Control and Estimation Using
  Nonlinear Programming: Second Edition}.
\newblock Advances in Design and Control, Society for Industrial and Applied
  Mathematics, 2010.

\bibitem{MartinThesis}
M.~P. Neuenhofen, {\em Quadratic integral penalty methods for numerical
  trajectory optimization}.
\newblock PhD thesis, Imperial College London, 2022.

\bibitem{TACDAIR}
Y.~Nie and E.~C. Kerrigan, ``Solving dynamic optimization problems to a
  specified accuracy: An alternating approach using integrated residuals,''
  {\em IEEE Transactions on Automatic Control}, vol.~68, no.~1, pp.~548--555,
  2023.

\bibitem{neuenhofen2023numerical}
M.~P. Neuenhofen, E.~C. Kerrigan, and Y.~Nie, ``Numerical comparison of
  collocation vs quadrature penalty methods,'' in {\em 2023 62nd IEEE
  Conference on Decision and Control (CDC)}, pp.~4285--4290, IEEE, 2023.

\bibitem{rao2010finite}
S.~S. Rao, {\em The Finite Element Method in Engineering}.
\newblock Elsevier Science, 2010.

\bibitem{9303897}
M.~P. Neuenhofen and E.~C. Kerrigan, ``A direct method for solving integral
  penalty transcriptions of optimal control problems,'' in {\em 2020 59th IEEE
  Conference on Decision and Control (CDC)}, pp.~4822--4823, 2020.

\bibitem{BBSOCP}
E.~R. Pager and A.~V. Rao, ``Method for solving bang-bang and singular optimal
  control problems using adaptive {R}adau collocation,'' {\em Computational
  Optimization and Applications}, vol.~81, p.~857–887, Apr. 2022.

\bibitem{ICLOCS2}
Y.~Nie, O.~Faqir, and E.~C. Kerrigan, ``Iclocs2: try this optimal control
  problem solver before you try the rest,'' in {\em 2018 UKACC 12th
  International Conference on Control (CONTROL)}, pp.~336--336, IEEE, 2018.

\bibitem{wachter2006implementation}
A.~W{\"a}chter and L.~T. Biegler, ``On the implementation of an interior-point
  filter line-search algorithm for large-scale nonlinear programming,'' {\em
  Mathematical Programming}, vol.~106, no.~1, pp.~25--57, 2006.

\bibitem{goddard1920method}
R.~H. Goddard, ``A method of reaching extreme altitudes,'' 1920.

\bibitem{tsien1951optimum}
H.~S. Tsien and R.~C. Evans, ``Optimum thrust programming for a sounding
  rocket,'' {\em Journal of the American Rocket Society}, vol.~21, no.~5,
  pp.~99--107, 1951.

\bibitem{TACRateConstraint}
Y.~Nie and E.~C. Kerrigan, ``Efficient implementation of rate constraints for
  nonlinear optimal control,'' {\em IEEE Transactions on Automatic Control},
  vol.~66, no.~1, pp.~329--334, 2021.

\bibitem{Ventilator2020}
E.~C. Kerrigan, Y.~Nie, O.~Faqir, C.~H. Kennedy, S.~A. Niederer, J.~A.
  Solis-Lemus, P.~Vincent, and S.~E. Williams, ``Direct transcription for
  dynamic optimization: A tutorial with a case study on dual-patient
  ventilation during the {COVID}-19 pandemic,'' in {\em 2020 59th IEEE
  Conference on Decision and Control (CDC)}, pp.~2597--2614, 2020.

\end{thebibliography}
\bibliographystyle{ieeetr}

\end{document}